\documentclass[12pt]{iopart}
\usepackage{graphicx}
\begin{document}

\title[The Contribution of the particle mass to the interaction measure]{Investigation on the contribution of the particle mass to the interaction measure}

\author{Mingmei Xu$^{1,2}$, Meiling Yu$^3$ and Yuanfang Wu$^{1,2}$}

\address{$^1$ Institute of Particle Physics, Central China Normal
University, Wuhan 430079, China}
\address{$^2$ Key Laboratory of
Quark \& Lepton Physics (Central China Normal University), Ministry
of Education, China}
\address{$^3$ Information engineering school, Hangzhou Dianzi
University, Hangzhou 310018, China}
\ead{xumm@iopp.ccnu.edu.cn}
\begin{abstract}
Collective phenomena from RHIC and LHC experiments indicate a
strongly coupled quark gluon plasma. Finite temperature lattice QCD
calculations show the interaction measure, $\Delta\equiv(\epsilon-3p)/T^4$, is sizeable over a
considerable range of temperatures above the deconfinement temperature,
which may also indicate that the plasma is strongly interacting. For the
ideal gas with massless particles, the interaction measure is zero. A
nonzero value is probably due to either the interaction or the mass. In
order to see the contribution of the particle mass to the interaction
measure, in this paper we study a system without any interactions,
i.e., an ideal gas with massive particles. After assembling the
standard formulas of the quantum statistics with relativistic energy, we
calculate the energy density, the pressure and the interaction measure. We find
that their dependences on temperature reproduce the qualitative
features of the lattice result. The interaction measure is nonzero for an
ideal gas, which demonstrates that the particle mass contributes to the interaction
measure. By our estimate, in the interaction measure obtained by the
 lattice calculation, quark mass contributes less than (40-50)\%. There are sizeable residue interactions in the deconfined phase.
\end{abstract}

\maketitle
\section{Introduction}
\def\tc{T_{\rm c}}
\def\d{{\rm d}}

A new state of matter, the quark gluon plasma (QGP), has been expected
since the 1970s of last century~\cite{qgp-early,qgp-early-2,wQGP}. Its experimental production in heavy ion
collisions is later well established by the evidences at
RHIC~\cite{RHIC-qgp-1,RHIC-qgp-2,RHIC-qgp-3,RHIC-qgp-4}. Many
discussions have been focused on the question whether this plasma
is weakly or strongly coupled~\cite{review-ws}.

The Hard Thermal Loop effective theory (HTL) is based on a physical
picture of the QGP as a gas of weakly coupled quasi-particles ---
quarks or gluons with temperature dependent effective masses and
couplings. It can well reproduce the thermodynamics from lattice data above $2\tc$, where $\tc$ is the deconfinement
temperature~\cite{HTL-1,HTL-2,HTL-3,HTL-4,HTL-5,HTL-6}. Even an ideal gas
with thermal gluon mass can describe the gluon lattice data for
thermodynamical quantities\cite{thermal-mass-gluon}. Thus, the
thermodynamics obtained by lattice QCD appears to be consistent with a weak coupling picture
of the QGP. However, the elliptic flow results at RHIC and the LHC
suggest a rapid thermalization and a very small value of the
viscosity-over-entropy-density ratio $\eta/s$, which is
inconsistent with the weak coupling calculations based
on kinetic theory~\cite{review-ws}. Such a small ratio can be obtained by the AdS/CFT
correspondence at
strong coupling\cite{AdS-1,AdS-2}. Thus, the small value of $\eta/s$ supports a strongly coupled QGP (sQGP).

It is intriguing that one system can be estimated as weakly coupled
for some phenomena while strongly coupled for others. It is
still conceivable since the coupling constant runs with the typical
momenta exchanged in the interactions on different space-time
scales~\cite{review-ws}.

The interaction measure, $\Delta\equiv(\epsilon-3p)/T^4$, defined in
terms of the energy density $\epsilon$ and the pressure $p$, is usually used
to measure the interaction between partons in the lattice QCD. For an
ideal gas with massless particles, energy density is equal to 3 times
pressure and $\Delta=0$. It is thought that sizeable $\Delta$ may
indicate strong interaction. The lattice QCD calculations indeed
find that $\Delta$ is sizeable at temperature
region $T=(1-2)\tc$~\cite{dlt-lattice}, and it appears to support a strong
coupling picture for the QGP. However, a non-interacting
quasi-particle description with only temperature-dependent effective
mass can also well explain the behavior of the interaction measure from
lattice~\cite{thermal-mass-gluon,effective-mass-01,effective-mass-02}.
The aim of the present study is to give another explanation on the
interaction measure obtained by lattice. We study a system without
any interactions, i.e., an ideal gas with massive particles. We first
calculate the energy density, the pressure and the $\Delta$. We find that their
dependences on temperature reproduce the qualitative features of the
lattice result. $\Delta$ is nonzero for an ideal
gas, which demonstrates that the particle mass contributes to the interaction measure. We then
further show that, in contrary to
\cite{thermal-mass-gluon,effective-mass-01,effective-mass-02}, in
our study the particle mass can not fully explain the interaction
measure obtained by lattice. There are sizeable residue interactions
in the deconfined phase. It may give us some new insight into the
nature of the plasma above the deconfined point.

\section{Thermodynamics of an ideal gas with massive particles}
Since both the interaction and the particle mass contribute to the
interaction measure, in this section we will study a system without any
interactions, i.e., an ideal gas with massive particles, to separate their contributions.  The reason we use an ideal gas is that we want to see the pure contribution of the particle mass when there are not any interactions.

For an ideal gas with particle mass $m$ at an equilibrium state
described by $(T,\mu,V)$, the standard formulas of the quantum
statistics give the energy density
\begin{equation}
\epsilon(T,\mu ;m)=g\int\frac{\d ^3 k}{(2\pi)^3}\varepsilon f_{\rm
F,B},
\end{equation}
and the pressure
\begin{equation}
p(T,\mu ;m)=\frac{g}{3}\int\frac{\d ^3
k}{(2\pi)^3}\frac{|\overrightarrow{k}|^2}{\varepsilon}f_{\rm F,B},
\end{equation}
where $g$ is the degeneracy factor, and $f_{\rm F,B}$ stands for the
single-particle distribution function with
\begin{eqnarray}
f_{\rm F}(\varepsilon,\mu,T)=\frac{1}{\e^{(\varepsilon-\mu)/T}+1}
\quad {\rm for\; fermions,}\\
f_{\rm B}(\varepsilon,\mu,T)=\frac{1}{\e^{(\varepsilon-\mu)/T}-1}
\quad {\rm for\; bosons.}
\end{eqnarray}
Using the relativistic dispersion relation
$\varepsilon=\sqrt{|\overrightarrow{k}|^2+m^2}$, we obtain
\begin{equation}
3p=g\int\frac{\d ^3
k}{(2\pi)^3}\frac{\varepsilon^2-m^2}{\varepsilon}f_{\rm
F,B}=g\int\frac{\d ^3
k}{(2\pi)^3}(\varepsilon-\frac{m^2}{\varepsilon})f_{\rm F,B}\leq
\epsilon.
\end{equation}
The above equation tells us that the presence of mass always reduces
the pressure below $\epsilon/3$. It is easy to understand because
massive particles move slowly at a given temperature and thus the
thermal pressure reduces. The interaction measure is then
\begin{equation}
\epsilon-3p=g\int\frac{\d ^3
k}{(2\pi)^3}\frac{m^2}{\varepsilon}f_{\rm F,B}.
\end{equation}

In the following we fix the chemical potential $\mu=0$ for
comparison with lattice results. At $\mu=0$, the pressure in equation (2),
the energy density in (1) and the interaction measure in (6) are only functions of
$T$ and $m$. In this paper the integrals for the quantum statistics are
done numerically.

Besides, we also do a Boltzmann statistics, i.e., the quantum
distributions are replaced by the Boltzmann distribution in the above
equations, which can be analytically done. For the pressure, a
technique of integrating by parts with the relation
$\frac{\partial}{\partial
p}\e^{-\varepsilon/T}=-\frac{1}{T}\frac{p}{\varepsilon}\e^{-\varepsilon/T}$
is used. Finally we obtain
\begin{eqnarray}
p(T;m)=\frac{g}{3}\int\frac{\d ^3
k}{(2\pi)^3}\frac{|\overrightarrow{k}|^2}{\varepsilon}\e^{-\varepsilon/T}
=\frac{T^4}{2\pi^2}x^2K_2(x),
\\
\epsilon(T;m)=T\frac{\d p}{\d T}-p(T)=\frac{3T^4}{2\pi^2}x^2
K_2(x)+\frac{T^4}{2\pi^2}x^3 K_1(x),\\
\epsilon-3p=\frac{T^4}{2\pi^2}x^3 K_1(x),
\end{eqnarray}
with $x=m/T$.

At fixed mass, e.g. $m=1.0$ GeV, $p(T)$, $\epsilon(T)$ and
$\Delta(T)$ are shown in figure 1(b), for both quantum statistics
and Boltzmann statistics. For easy of comparing, the lattice results
are shown in figure 1(a). The lattice studies the equation of state
of a real QCD matter at $\mu=0$. In figure 1(a), a continuous
increase of the energy density and the pressure is seen, indicating a
crossover between the hadron gas and the quark gluon plasma, which is
confirmed by later reference~\cite{crossover}. There are more recent
results~\cite{hotQCD,Wuppertal-Budapest,hotQCD2} on the QCD equations of state
 than shown in figure 1(a) from reference~\cite{dlt-lattice}. In the temperature
region $\tc<T<2\tc$ ($\tc=170$MeV), the interaction measure, denoted as
$I$ in figure 1(a), is sizeable and does not vanish.

\begin{figure}[t]
\centering
\includegraphics[width=2.6in]{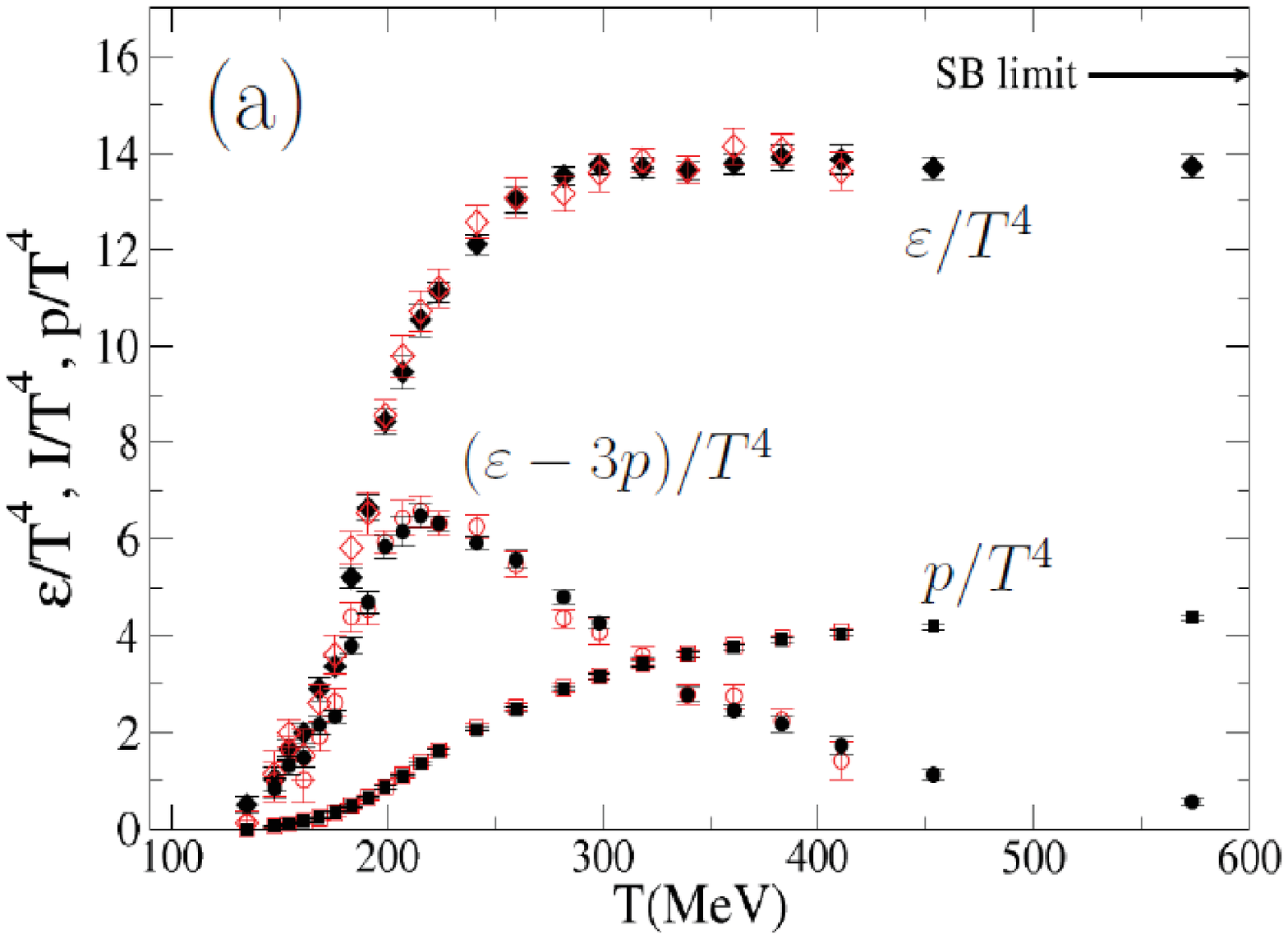}\hskip 0.5cm
\includegraphics[width=2.6in]{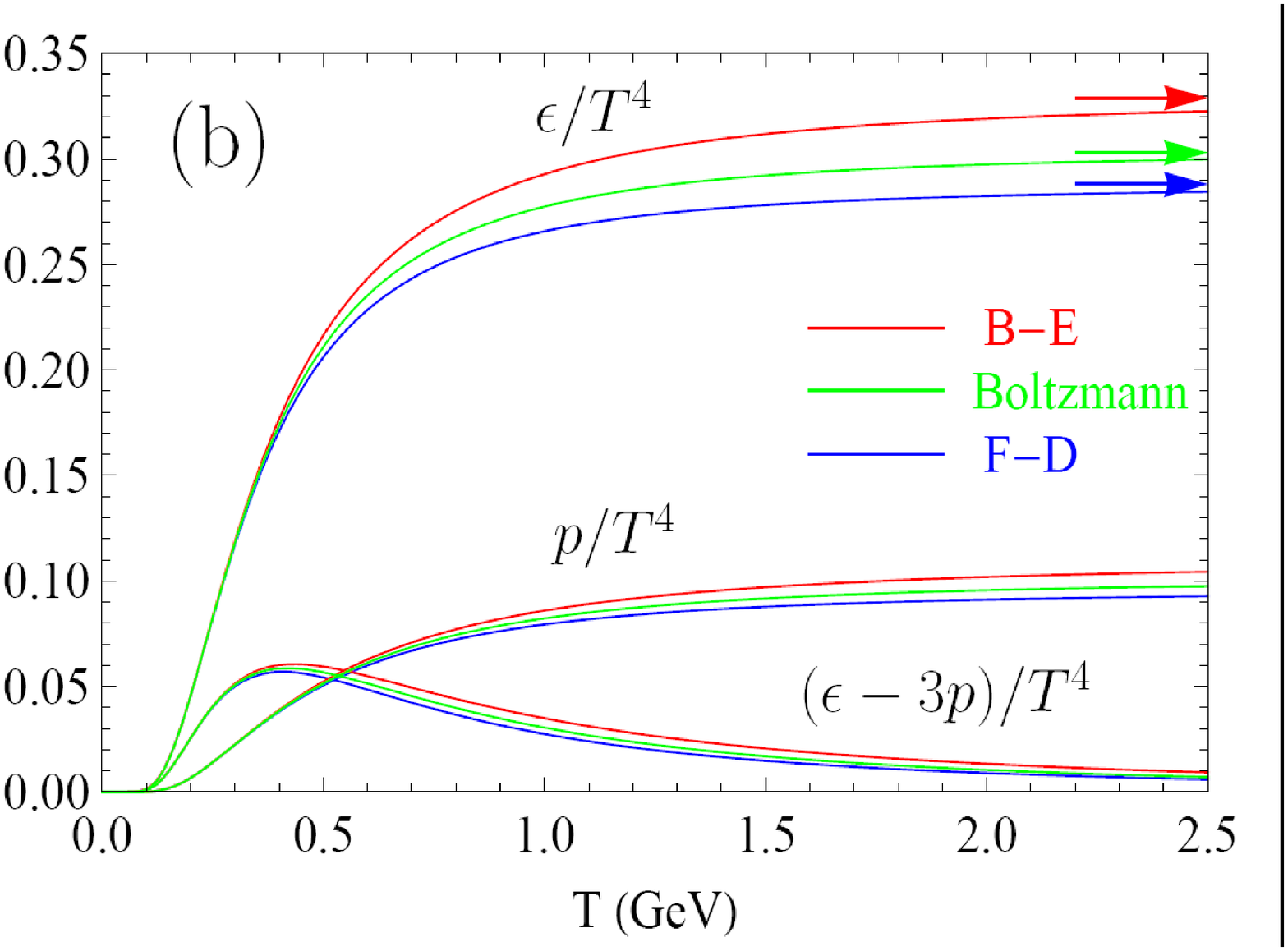}
\caption{Energy density, pressure and interaction measure are
plotted. (a) the lattice QCD result of strongly interacting matter
from figure 7 in reference~\cite{dlt-lattice} with $m_{\rm
ud}\approx 0.1 m_{\rm s}$. Diamonds, squares and circles are for
energy density, pressure and interaction measure, respectively.
(Reprinted figure with permission from Bernard C et al, Phys. Rev.
D, 75, 094505, 2007. Copyright (2007) by the American Physical
Society.) (b) the result of an ideal gas with particle mass
$m=1.0$GeV. Red lines for bosons, blue lines for fermions and green
lines for Boltzmann approximation. The arrows near the right axis
indicate the corresponding Stefan-Boltzmann limits.(Color online)}
\end{figure}

The analysis is as follow:
\begin{enumerate}
  \item In figure 1(b), $\epsilon/T^4$ and $p/T^4$ are increasing with $T$
   and saturate at high temperature. At high temperature limit, $p$ and $\epsilon$ approach Stefan-Boltzmann limits
marked as arrows in the figure. (The SB limits are given in Appendix A.) $\Delta/T^4$ is nonzero and
   peaks at the temperature region where the pressure and the energy density
    increase rapidly, which is similar to the lattice result.
     It is interesting that, even an ideal gas has the similar trend as the
     lattice results for the
  QGP thermodynamics. Besides, another kind of a more complicated
  weak-coupling calculations from the kinetic theory\cite{HTL-1,HTL-2,HTL-3,HTL-4,HTL-5,HTL-6} can describe
  the lattice data not only the trend but also quantitatively. In addition, an ideal gas with temperature-dependent mass also leads
   to good agreement with lattice~\cite{thermal-mass-gluon,effective-mass-01,effective-mass-02}.
   The system used here is quite simple, which only reproduce the qualitative features of
   thermodynamic quantities.
   The aim of this paper is to investigate how much the particle mass contributes to the interaction measure. By comparing our results with the lattice results, we can obtain the pure contribution of the particle mass.
    A quantitative analysis will be given in the subsequent paragraphs.
  \item As discussed before, for an ideal gas with massless particles, $\Delta=0$. In figure 1(b), it can be seen that $\Delta$ is nonzero for an ideal gas with massive particles. The only difference between the two systems is the particle mass. Thus, the lesson we learn from here is
that, a nonzero $\Delta$ does not certainly mean strong coupling
since the particle mass contributes to this quantity.
\item Here we use $m=1.0$ GeV as an example to show the behavior of the interaction measure. Similar to the lattice result, there is a peak at the temperature region where the energy density and the pressure grow fast. By changing the particle mass, the curve of
the interaction measure will shift its position, as shown in figure
2. The $\Delta$ peak
shifts to low $T$ for small mass and to high $T$ for large mass. And the peak height is independent on mass for an
ideal gas. This statement is valid for an ideal gas with either quantum statistics or Boltzmann
statistics.
  \item In figure 1(b), the peak height of $\Delta/T^4$ for the ideal gas is
  0.057 (this value is for Fermi-Dirac statistics, 0.059 for
  Boltzmann statistics, 0.061 for Bose-Einstein statistics),
  which is much smaller
  than the lattice calculation. In figure 1(a), the peak height is about 6.5, while
  newer results from HotQCD collaboration favor a value of about 5.0 for
  the peak height~\cite{hotQCD}, and the continuum extrapolated result of the
  Wuppertal-Budapest collaboration indicates a peak height of about 4.1~\cite{Wuppertal-Budapest}.
   The great
  disparity in peak height between figure 1(a) and figure 1(b) can be partly attributed to the number of degrees of
  freedom of constituent particles. The degeneracy factor $g=1$ in the ideal gas system in figure 1(b).
  However, $g$ in the QCD matter considered by lattice refers to a large number of degrees of freedom, including
   gluon and three flavor quark, anti-quark which further count
    color, spin degeneracy.

\begin{figure}[t]
\centering
\includegraphics[width=2.8in]{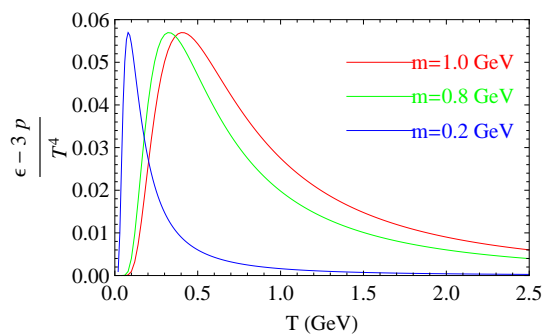}
\caption{The mass dependence of the interaction measure. Red, green, blue
lines are for mass 1.0, 0.8, 0.2 GeV respectively. This plot is
calculated from Fermi-Dirac statistics.(Color online)}
\end{figure}

Taking the pressure for an example. In
an ideal multicomponent gas, the partial pressure of one component
is $d\times p_{m}$, where the particle degeneracy factor $d$
represents the number of the internal degrees of freedom for the
particle with mass $m$, $p_{m}$ represents the pressure without
any internal degrees of freedom. The total pressure of an ideal
multicomponent gas can be expressed as
\begin{equation}
p(T;m_1,\ldots,m_n)=d_1 p(T,m_1)+\ldots+d_n p(T,m_n).
\end{equation}
Similarly,
\begin{equation}
\epsilon(T;m_1,\ldots,m_n)=d_1 \epsilon(T,m_1)+\ldots+d_n
\epsilon(T,m_n),
\end{equation}
and the interaction measure is
\begin{equation}
\Delta(T;m_1,\ldots,m_n)=d_1 \Delta(T,m_1)+\ldots+d_n \Delta(T,m_n).
\end{equation}
In lattice calculations, only quark mass contributes to the interaction measure since gluon is massless, according to equation (6).
For three flavor QCD,
there are two masses: $m_{\rm ud}$ and $m_{\rm s}$. $\Delta$ for these two
masses are peaks at different positions. According to (12),
$\Delta$ for the mixed gas will be a double-peak curve.

The maximum value of $\Delta$ in equation (12), denoting as
$\Delta^{\rm max}$, is less than
 \begin{equation}
 \Delta^{\rm max}<d_1 \Delta^{\rm max}_{m_1}+\ldots+d_n \Delta^{\rm max}_{m_n},
\end{equation}
where $\Delta^{\rm max}_{m}$ represents the peak height of the $\Delta$
curve with fixed mass $m$. Since the peak height is independent on mass for an ideal gas, then $\Delta^{\rm max}_{m_1}=\Delta^{\rm
max}_{m_n}=\Delta_0=0.057$. Thus we can infer that
\begin{equation}
\Delta^{\rm max}<d_{\rm total}\Delta_0,
\end{equation}
where $d_{\rm total}=d_q+d_{\overline{q}}=36$ for three flavor QCD,
with the quark degeneracy
$d_q=d_{\overline{q}}=N_cN_sN_f=18$~\cite{book-Wong}. Finally we get
    \begin{equation}
    \Delta^{\rm max}<d_{\rm total}\Delta_0= 2.1.
    \end{equation}
The value at the right hand side of the above equation is about 40\% of the lattice
result 5.0 reported by HotQCD Collaboration~\cite{hotQCD}, and about 50\% of the
lattice result 4.1 reported by Wuppertal-Budapest Collaboration~\cite{Wuppertal-Budapest}.
Therefore, from our analysis we can conclude that the pure contribution of the particle mass is not larger than (40-50)\%.

In this paper, we only focus on the peak height which depends on the interaction. The peak position may also shift by additional interaction. Thus the peak position, i.e., the temperature where the interaction measure gets maximum, for an ideal gas can not be used to compare with that of the lattice QCD.
\end{enumerate}

\section{Conclusions}

We use an ideal gas with massive particles to calculate the energy density, the pressure and the interaction measure, and compare them with that of the lattice results. An ideal gas has the similar trend as the lattice results for the QGP thermodynamics. We reproduce the qualitative features of the lattice results on the energy density, the pressure and the interaction
measure by an ideal gas with massive
particles. The interaction measure is nonzero for an ideal gas, which demonstrates that a nonzero $\Delta$ does not certainly mean strong coupling.
Particle mass can make the interaction measure nonzero. The $\Delta$
measured in the lattice QCD includes the contribution from the particle mass
because they indeed use quark mass in the lattice
calculation~\cite{dlt-lattice,quarkmass}. After counting the
degeneracy number of the QGP, $\Delta$ contributed by the particle mass
explains less than (40-50)\% of the lattice result. We infer that the
other contribution comes from interaction. That
means nonzero $\Delta$ from lattice calculation does not mean strong
coupling but does not exclude a strong coupling picture of QGP. It
is still hard to know how strong is the coupling since it is hard to
know the relation of the interaction measure with the coupling. The question how strong is the coupling needs further study.

\ack

We thank Na Li for helpful discussions. This work is supported by the NSFC of china with project nos 11005045, 10835005,
11047124 and by CCNU-QLPL Innovation Fund (QLPL2011P01).

\appendix\section{Stefan-Boltzmann limits}

Stefan-Boltzmann limit is the value for a non-interacting gas with
massless particles.

For Boltzmann particles, since $x^2 K_2(x)\rightarrow 2$ for
$x\rightarrow 0$, equation (7)(8) becomes
\begin{eqnarray}
p(T;m)=\frac{T^4}{2\pi^2}x^2K_2(x)\rightarrow\frac{T^4}{\pi^2},
\\
\epsilon(T;m)=T\frac{\d p}{\d T}-p(T)\rightarrow\frac{3T^4}{\pi^2}.
\end{eqnarray}
Therefore,
\begin{eqnarray}
\frac{\epsilon}{T^4}\rightarrow\frac{3}{\pi^2}\approx 0.303\quad
{\rm for\; Boltzmann\; particles,}
\end{eqnarray}
as marked in figure 1(b).

For quantum particles, ref~\cite{book-Wong} gives
\begin{eqnarray}
\frac{\epsilon}{T^4}\rightarrow\frac{7}{8}\frac{\pi^2}{30}\approx
0.288\quad {\rm for\; fermions,}\\
\frac{\epsilon}{T^4}\rightarrow\frac{\pi^2}{30}\approx 0.329\quad
{\rm for\; bosons,}
\end{eqnarray}
which are also marked in figure 1(b).

\def\ea{{\it et al.}}

\end{document}